\documentclass[twocolumn,%floatfix,
showpacs,prd,aps,eqsecnum,tightenlines,superscriptaddress]{revtex4}
\usepackage{graphicx}
\usepackage{dcolumn}% Align table columns on decimal point
\usepackage{bm}% bold math

\def\mm{-\,-}
\def\pp{++}

\begin{document}

%\preprint{UTBRG-2003-001, astro-ph/03mmnnn}

\title{The coalescence remnant of spinning binary black holes}

\author{J. Baker} \affiliation{Laboratory for High Energy
Astrophysics,  NASA Goddard Space Flight Center, Greenbelt, Maryland
20771}
\author{M. Campanelli} \affiliation{Center of Gravitational Wave Astronomy
and Department of Physics and
Astronomy,  The University of Texas at Brownsville, Brownsville, Texas
78520}
\author{C. O. Lousto} \affiliation{Center of Gravitational Wave Astronomy
and Department of Physics and
Astronomy,  The University of Texas at Brownsville, Brownsville, Texas
78520}
\author{R. Takahashi} \affiliation{Theoretical Astrophysics
Center, Dk-2100 K\o benhavn \O , Denmark}

\date{\today}

\begin{abstract}
%{\bf We should write a brand new abstract here}
We compute the gravitational radiation generated in the evolution of a
family of close binary black hole configurations, using a combination
of numerical and perturbative approximation methods.  We evolve the
binaries with spins, $s$, aligned or counter-aligned with the orbital
angular momentum from near the innermost stable circular orbit (ISCO)
down to the final single rotating black hole.  For the moderately
spinning holes studied here the remnant Kerr black hole formed at the
end of an inspiral process have a rotation parameter
$a/M\approx0.72+0.32(s/m_H)$, suggesting it is difficult (though not
excluded) to end up with near maximally rotating holes from such
scenarios.
\end{abstract}

\pacs{04.25.Dm, 04.25.Nx, 04.30.Db, 04.70.Bw} \maketitle

\section{Introduction}\label{Sec:Intro}

The coalescence of two black holes of comparable size will provide the
most extreme dynamical tests of General Relativity's predictions. Such
coalescences are an anticipated outcome of galactic collision and core
merger. The first galaxy with binary active galactic nuclei has, recently,
been discovered by X-ray observations of NGC 6240 with the Chandra
Observatory \cite{Komossa:2002tn}. Possible evidence of merger events
has also been recently presented in radio observations of X-shaped jet
morphologies, which may have been produced by a sudden change in the
central black hole's spin axis caused by a supermassive black hole -
black hole merger \cite{Merritt:2002hc}.

We focus on modeling the final moments of a binary black hole merger,
which generates the last few cycles of radiation, carrying away a
significant fraction of the system's energy and angular momentum.  In
a previous paper\cite{Baker:2002qf} we have studied non-spinning
binary black holes.  Our goal here is to extend this treatment to deal
with moderately spinning holes focusing, in particular on the
characteristics of the final remnant black hole which the coalescence
produces.  While we generically expect collisions of binary systems
with significant mass ratios, at a given mass systems with a mass
ratio near unity are expected to produce the strongest gravitational
raditation, producing the the largest effect on the state of the final
black hole. We will restrict to this case here.  As we are interested
in the dynamics of these systems at their strongest (most non-linear)
moment, we are impelled to apply a model which includes numerical
treatment of Einstein's equations in their fully nonlinear form. The
results here are thus complementary to previous studies in both the
slow-motion weak-field post-Newtonian (PN) approximation, appropriate
when the black holes are still far
apart\cite{Damour:2001tu,Blanchet:2002fk} and extreme mass ratio
treatments based on near-geodesic
motion\cite{Saijo:1998mn,Hughes:2002ei}.

In this paper, we adopt the viewpoint that the individual black hole
spins vary only slowly as the system approaches the ISCO.  In general
then, we expect pairings of black holes near ISCO with arbitrary
spins.  We treat the special case of spins either aligned or
anti-aligned with the orbital angular momentum paying particular
attention to how this addition of spin to the problem affects
resulting angular momentum of the finally produced black hole. These
configurations exclude any precessional effects of the strong field
spin interactions, but include the case expected to produce the most
rapidly rotating remnant.  We apply the
Lazarus~\cite{Baker00b,Baker:2001sf,Baker:2001nu,Baker:2002qf},
approach to treat several systems with moderate spin, allowing us to
calculate the radiative loss of energy and angular momentum.

\section{The model}\label{Sec:model}

The goal of the Lazarus approach is to exploit a broad range of
analytic and numerical techniques to model approximately black hole binaries
during the different stages of the coalescence.  In particular, while
the Close Limit (CL) approximation Ref.\ \cite{Price94a,Lousto99a} can
describe the ring-down of the finally formed black hole, Numerical
Relativity (NR) simulations are needed in order to provide a
description of the system in the strong non-linear merger stage of the
collision.  We also need a description for the system applicable in
the Far Limit (FL) which can provide initial values for the NR
simulations.  In this paper we will apply the same CL and NR
treatments which we used to study the nonspinning case
\cite{Baker00b,Baker:2001sf,Baker:2001nu,Baker:2002qf}, but we have extended
the FL model to allow spinning black holes.  

\subsection{Initial data}

We handle spinning black holes in the nearly same way as for the
nonspinning case, with the Bowen-York-puncture ansatz
\cite{Brandt97b}.  To study plunge radiation we want to begin our
simulations with black holes at the innermost stable quasi-circular
orbit (ISCO) for {\it spinning} binary black holes as determined for
the Bowen-York-puncture initial data.  Various approaches have been
developed to compute the location and frequency of the ISCO. We will
use results based on the effective potential method of \cite{Cook94}
as generalized by Pfeiffer, Teukolsky, and Cook \cite{Pfeiffer:2000um}
for spinning holes.  A sequence of quasi-circular orbit configurations
determined by minimizing the binding energy with respect to separation
along sequences of constant mass ratio, orbital angular momentum and
spin.  The ISCO is the limit point of this sequence.  
We also note that the use of Refs.~\cite{Cook94,Pfeiffer:2000um}
results, strictly valid for the "image method", holds for the
"punctures" since the difference outside the horizons, are very small
\cite{Abrahams95c} in practice for the fairly detached black holes
studied here.

The parameters of seven ISCO orbits are given in Table \ref{SItable}.
The black
holes have parallel spins aligned and counter-aligned with the orbital
angular momentum as in Ref.\ \cite{Pfeiffer:2000um}.  We label the
different cases by $S$, the $z$-component of each individual black
hole's spin, scaled by the square of its mass.  This corresponds to
the scaled angular momentum parameter $a/m$ for each hole. In the
tables, $\ell$ represents the proper distance between throats, and
$E_b$ is the binding energy for the given configuration.  $Y$ is the
coordinate location of the punctures in conformal space (on the
$y$-axis), $P$ is the linear momentum of the holes (as measured from
infinity) and are chosen opposite and transversal to the line joining
the holes in conformal space, so that for our simulations the total
angular momentum, J, is in the $z$-direction.  We denote by $s$ the
individual spin of the holes, and $L$ the orbital angular momentum of
the system.  The angular frequency of the quasi-circular orbit,
$\Omega$, is determined from $dE_b/dL$. Finally, $m$ denotes the
individual 'puncture' (or bare) masses of the holes, and $E_b$ is the
binding energy for the given configuration.  All quantities are
normalized to $M$, the total ADM mass of the system, and are computed
on the initial time slice.

%Note that the Bowen-York initial data family does not have the Kerr
%limit for larger separations, as the Kerr metric does not have
%conformally flat slices \cite{Garat:2000pn}. Thus, there will be some
%extra `spurious' radiation content introduced by the spins.  In the
%far limit, these data describe two distorted Kerr black holes rather
%that the expected Kerr solutions.  Fortunately, for the moderate
%spins, aligned and anti-aligned with the orbital angular momentum,
%studied here, this effect can be estimated to be contained below $1\%$
%of the radiated energy during the plunge \cite{Dain:2002ee}.  Thus, to
%the level of current accuracy, our results are not significantly
%affected byt his effect.

\begin{table}
%\begin{widetext}
\caption{Spin - ISCO data based on the effective potential method
applied to Bowen-York data}
\begin{ruledtabular}
\begin{tabular}{llllllll}\label{SItable}
 $S$ & \mm0.50 & \mm0.37 & \mm0.25 & \mm0.12 &\pp0.00&\pp0.08&\pp0.17\\
 \hline
 $\ell/M$& 7.16 & 6.80 &6.29 &5.70 &5.05 &4.70 &4.04 \\
 $\pm Y/M$& 1.930  &1.831  &1.651  &1.433  &1.193  &1.062  &0.810  \\
 $\pm P/M$&0.230 &0.236 &0.254 &0.282 &0.326 &0.356 &0.451 \\
 $s/M^2$&-0.129 &-0.0958 &-0.0649 &-0.0313 &0.00 & 0.021 & 0.0449 \\
 $J/M^2$&
% $J\footnote{Total: Orbital plus spin}/M^2$&
 0.629 &0.672 &0.709 &0.747 &0.779 &0.799 &0.820 \\
 $L/M^2$&
% $L\footnote{due to orbital motion only}/M^2$&
 0.887  &0.863 &0.838 & 0.809 &0.779 &0.757 &0.730  \\
 $M\Omega$&0.098 &0.105 &0.118 &0.136 &0.162 &0.182 &0.229  \\
 $m/M$& 0.424& 0.448&    0.46&     0.464&     0.455&     0.45&     0.435\\
 $E_b/M$ & -.0159 & -.0175 & -.0189 & -.0208 & -.023 & -.0250 & -.0279
% $E_b/M$ & -.01595 & -.01748 & -.01893 & -.02080 & -.02304 & -.02498 & -.02793
\end{tabular}
\end{ruledtabular}
%\end{widetext} 
\end{table}

%%%%%%%%%%%%%%%%%%%%%%%%%%%%%%%%%%%%%%%%%%%%%%%%%%%%%%%%%%%%%%%%%%%%%%%%%%

\section{
%Radiation from spinning black holes
Results and Discussion}\label{Sec:radiation}

In this section we present the results of our simulations beginning
from the data described in the last section.  As described in Ref.
\cite{Baker:2002qf}, we begin by numerically solving Einstein's equations.
The difficulty of performing these numerical simulations limits us to
brief evolutions, thus we next apply several criteria in determining
when, or whether, the numerical simulation has 'linearized' and
close-limit treatment should be applicable.  We first look at the
values of the $S$-invariant~\cite{Baker00a}, which we expect to be
within a factor of two of its background value, unity.  We also look
for independence in the waveforms on the time $T$ when we transition
from the numerical simulation to the close limit treatment.  In
particular, we expect to find independence of the waveform phase over
the most significant part of the waveform, as we vary $T$.  We also
expect to see some leveling of the dependence of the total radiation
energy as we vary $T$, though this tends to be more difficult to
achieve.  For each case we will determine waveforms, radiation energy
and angular momentum, and the state of the final black hole.

%\subsection{Aligned black holes}
%\subsubsection{$SI$++0.17}

\begin{figure}
\begin{center}
\begin{tabular}{@{}lr@{}}
\includegraphics[width=1.75in]{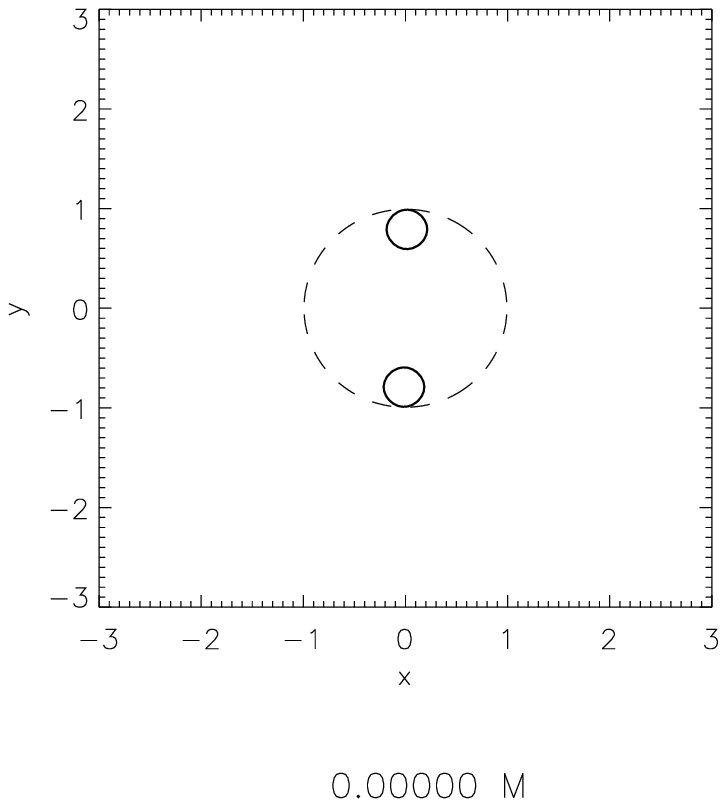} &
\includegraphics[width=1.75in]{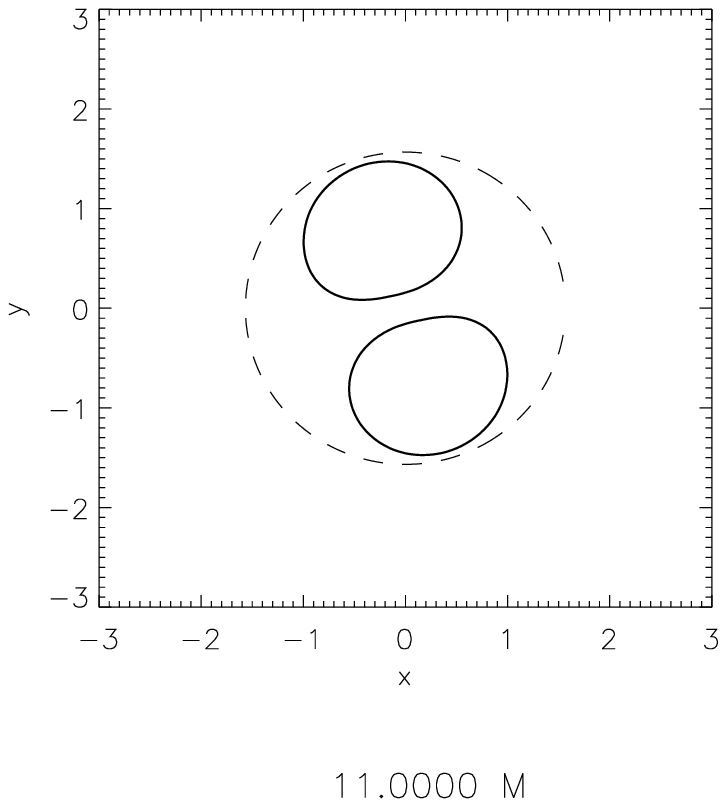}\\
\includegraphics[width=1.75in]{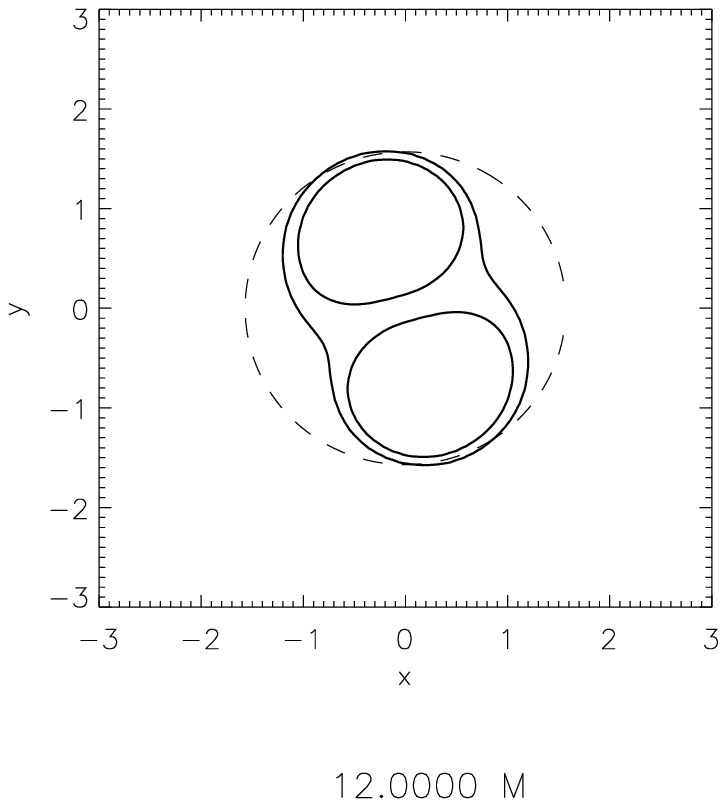} &
\includegraphics[width=1.75in]{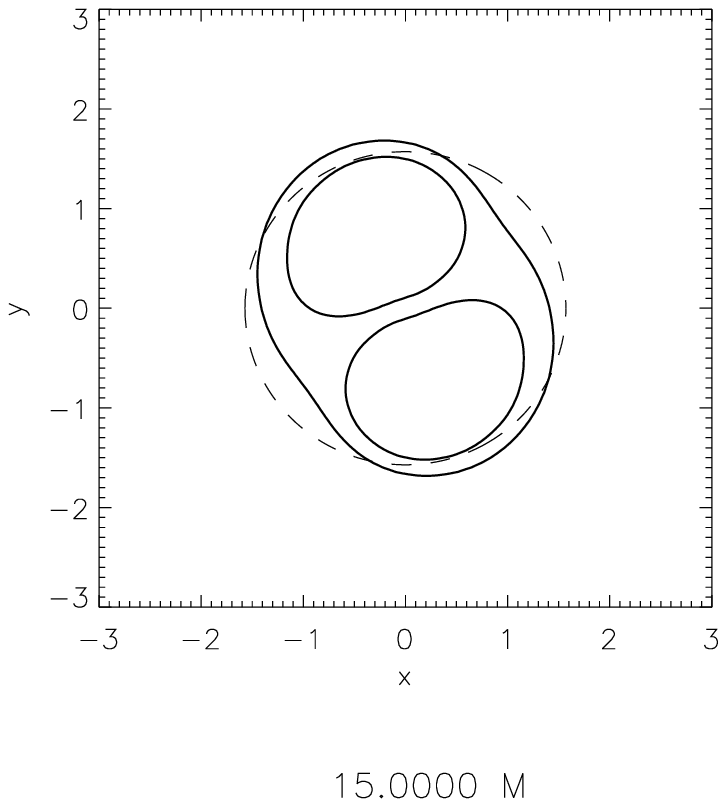}
\end{tabular}
\end{center}
\caption{Horizon formation for $SI$++0.17.  The solid curves show the 
numerically determined location of trapped surfaces including the
common apparent horizon of the final black hole.  The dashed curve
indicates the independently calculated horizon location of the
``background'' perturbed Kerr black hole needed in the CL part of the
treatment. }
\label{fig:AHSI++0.17}
\end{figure}

The first configuration we discuss, $SI$++0.17, is the case with the
strongest spins aligned with the orbital angular momentum.  

In this
case we are able to run the numerical simulation long enough to
numerically locate a common apparent horizon around the black holes.
Fig.\
\ref{fig:AHSI++0.17} shows four snapshots of the apparent horizon
surfaces~\cite{Alcubierre98b} on the orbital plane for the $SI$++0.17
case.  The dashed line shows the location of the apparent horizon of
the Kerr black hole, which we have identified as the background for
the Close Limit calculation.  The plots show that the two black holes
are well detached at $T=0$.  The `grid stretching' effect (due to the
vanishing shift we used during full numerical evolution) makes them to
grow in the coordinate space.  These plots are particularly useful to
extract qualitative information about the system.  Soon after a common
apparent horizon covers the system it tends to have an increasingly
spherical shape. The time of formation of a common apparent horizon
roughly gives an upper limit to the time at which linear theory can
take over as has been discussed in Ref.~\cite{Baker:2001sf}.  It is
expected that a common {\it event} horizon should have appeared some
$M$'s of time earlier in the evolution of the system.  On the independent 
basis of S-invariant and waveform robustness criteria, 
we estimate the linearization time to come at $T=7-9M$.  This is 
consistent with the early formation of a apparent horizon. 
The key
physical feature which actually makes the close limit approximation
effective is that the black holes share a common potential barrier
which appears even earlier than the common event horizon. Near the
time of linearization we estimate the location ``background horizon''
at $r_H\approx1.55$ in the numerical coordinates.

Continuing the evolution in the CL treatment, and after
two iterations of the background mass and the radiated
angular momentum (starting from the initial data values),
we estimate the total 
energy lost to radiation to be $1.7\%$ - $1.9\%$ of the system's initial 
energy.  Likewise, $0.06-0.08M^2$ of the initial angular momentum is lost,
resulting in a final Kerr black hole formed after
the plunge with $M_f\approx0.982M$ and $a_f/M_f\approx0.778$.

\begin{table}
\caption{Summary of results for spinning holes}
\begin{ruledtabular}
\begin{tabular}{llllll}\label{tutto}
$S$ & ++0.17 & ++0.08 & ++0.00 & -\,-0.12 & -\,-0.25 \\
\hline
$T/M$ & 7-9 & 9-11 & 9-11 & 10-12 & 11-12\\
$-\dot{E}/M(\%)$ & 1.7-1.9 & 2.3-2.5 & 2.4-2.6 & 1.9-2.1 & 1.9-2.1\\
$-\dot{J}/M^2$ & 0.06-0.08 & 0.09-0.10 & 0.09-0.10 & 0.09-0.10 & 0.09-0.10\\
$M_f/M$ & 0.982 & 0.976 & 0.975 & 0.980 & 0.980 \\
$a_f/M_f$ & 0.778 & 0.739 & 0.720 & 0.679 & 0.639\\
$J_i/M^2_i$ & 0.820 & 0.799 & 0.779 & 0.747 & 0.709 \\
$M\omega_{peak}$ & 0.50-0.55 & 0.45-0.50 & 0.45-0.50 & 0.40-0.50 & 0.40-0.50\\
$M\omega_{QN}$ &    0.573  &    0.552  &   0.542  & 0.524 & 0.508\\
$M/\tau_{QN}$   &  0.07714  &  0.07925  &  0.08006  & 0.08160 & 0.08285\\
$T_{CAH}/M$ & 12 & $> 15$ & $>17$ & -- -- & -- -- \\
\end{tabular}
\end{ruledtabular}
\end{table}

The attractive nature of the spin obrit coupling makes the $SI$++0.17,
with the strongest aligned component of spin angular momentum the
highest frequency and closest configuration at ISCO. As we decrease
$s$, we expect the larger separations to require larger linearization
times $T$.  This expectation is consistent with our calculations which
indicate a linearization time of $T=9-11M$ for the $SI$++0.08 and
$SI$++0.00 cases, and respectively $T=10-12M$ and $T=11-12M$ for the
antialigned spin cases $SI$--0.12 and $SI$--0.25. The summarized
results are shown in Table\ \ref{tutto}.  Overall, energy lost to
radiation is near $2\%$, but peaked at spinless case.  Although the
interaction time increases as we decrease $s$, suggesting more energy
could be lost here, the mean frequency of the raditation decreases so
that the energy is lost at a reduced rate for the antialigned spin
cases.  The decrease in frequency does not affect the rate at which
angular momentum can be radiated as dramatically, which seems to
results in a nearly flat $s$-dependence for the total loos of angular
momentum.  The ranges given in the table, are estimated from
self-consistency tests based on comparisons of perturbative and
numerical treatments.  Our results may also be subject to systematic
errors resulting from effects which may have led us to underestimate
the linearization time.  We would generally expect any such effect to
lead to an increase amount of energy and angular momentum radiated,
and a reduction in the final $a/m$.  The results are also subject to
errors in the initial model representing black holes at ISCO.  A more
detailed study of the non-spinning case in \cite{Baker:2002qf}
indicates some insensitivity in the radiation result to the precise
location of ISCO, but further dynamical testing of this class of
initial data should be carried out as advances in evolution techniques
make it possible.

\begin{figure}
\begin{center}
\includegraphics[width=3.2in]{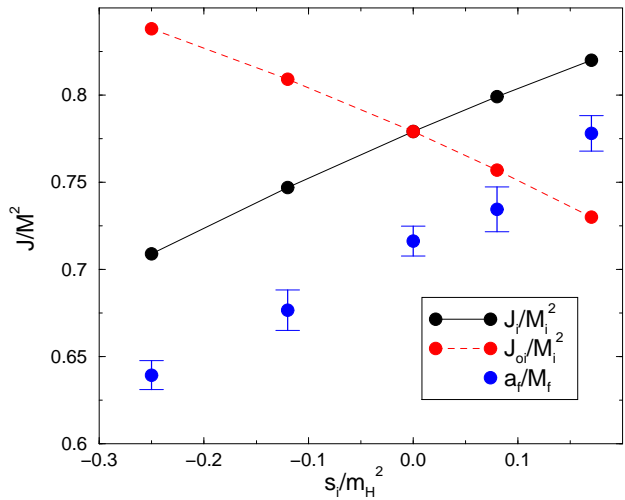}
\end{center}
\caption{Rotation parameter for the final remnant Kerr black hole form in 
coalescence of spining black holes. The plot shows dependence on the initial individual component of spin aligned with the orbital anglar momentum, and 
normalized to the horizon mass $s/m_H^2$. The initial obrital and total angular momenta $J_o$ and $J_i$ are also shown.}\label{fig:aM}
\end{figure}

A particularly interesting quantity derived from each of these results
is the angular momentum parameter for the final remnant Kerr holes
formed as a product of coalescence. The remnant black holes 
have a larger rotation parameter $a/M$, for the aligned spin cases than
for the anti-aligned ones. In Fig.~\ref{fig:aM} error bars are estimated by
taking the absolute maximum and minimum of observed the damped oscillations of
the energy and angular momentum radiated versus the transition time.
This provides us with a much larger error than self-consistency tests
suggest, but might be more
representative of the possible systematic errors of our approach.

It is interesting to note here that the rotation parameters of the
final Kerr hole $a_f/M_f$, of table~\ref{tutto} and Fig.~\ref{fig:aM}
are large, but still far from the maximally rotating hole, suggesting
it is hard to generate near maximally rotating single holes after the
plunge of two inspiralling holes when the have moderate individual
rotation parameters like the ones studied in this paper.  A curve
fitting to the values in the Fig.~\ref{fig:aM} gives
$J/M^2\sim0.779+0.2566(s/m_H^2)-0.0941(s/m_H^2)^2$ for the initial data
and $a/M\sim0.719+0.324(s/m_H^2)$ for the final Kerr
hole, where $m_H$ is the horizon mass of the individual holes and $s$
its individual spin. If we extrapolate the trend linearly toward
$s/m_H^2\sim1$, the result suggests we would require initially aligned
holes with spins $>0.85$ in order to approach a maximally rotating
remnant.  Our studies thus enhance the conclusions based on
complementary studies of binaries in the small mass ratio
limit\cite{Hughes:2002ei} which suggested that it is hard to form a
maximally rotating black hole by binary merger unless the mass ratio
is near unity.  Even in the equal mass case it would seem to take near
maximally spinning black holes to produce a maximally spinning
remnant.  Of course, on a broader astrophysical scenario,
rapidly rotating black holes may still be produced by accretion.

We finally note that we also report the binding energy $E_b$ of the initial
configurations in table \ref{SItable} for the nonspinning holes and in
table~\ref{tutto} for the binary spinning holes. If we interpret this
binding energy as a measure of the energy radiated in the form of
gravitational waves during the whole inspiral period until we reach the
initial configuration we use to estimate the plunge radiation, we came
to the notable conclusion that approximately as much energy is radiated
in the few cycles following the plunge as in the whole prior lifetime of the
binary.

\acknowledgments 
We thank H.Pfeiffer for making available unpublished data related to
Ref. \cite{Pfeiffer:2000um}. J.B. is supported
by the National Research Council.  C.O.L and M.C. have financial
support from Grants NSF-PHY-0140326 'Kudu' and NASA-URC-Brownsville.
The full nonlinear numerical work has been performed in LRZ (Germany)
and NERSC (under Contract No. DE-AC03-76SF00098).
\bibliographystyle{apsrev}
\bibliography{bibtex/references}
\thebibliography{spin_br}

\end{document}